\documentclass[twocolumn,english,aps,prl, twocolumn, superscriptaddress]{revtex4}
\usepackage[T1]{fontenc}
\usepackage[latin9]{inputenc}
\usepackage{longtable}
\usepackage{graphicx}

\makeatletter

\providecommand{\tabularnewline}{\\}

\@ifundefined{textcolor}{}
{%
 \definecolor{BLACK}{gray}{0}
 \definecolor{WHITE}{gray}{1}
 \definecolor{RED}{rgb}{1,0,0}
 \definecolor{GREEN}{rgb}{0,1,0}
 \definecolor{BLUE}{rgb}{0,0,1}
 \definecolor{CYAN}{cmyk}{1,0,0,0}
 \definecolor{MAGENTA}{cmyk}{0,1,0,0}
 \definecolor{YELLOW}{cmyk}{0,0,1,0}
 }

\makeatother

\usepackage{babel}

\begin{document}

\title{Fast Reconnection in High-Lundquist-Number Plasmas Due to the Plasmoid
Instability }

\author{A. Bhattacharjee }

\affiliation{University of New Hampshire, Durham, NH 03824 }

\affiliation{Center for Integrated Computation and Analysis of Reconnection and
Turbulence}

\author{Yi-Min Huang }

\affiliation{University of New Hampshire, Durham, NH 03824 }

\affiliation{Center for Integrated Computation and Analysis of Reconnection and
Turbulence}

\author{Hongang Yang }

\affiliation{Dartmouth College, Hanover, NH 03825}

\affiliation{Center for Integrated Computation and Analysis of Reconnection and
Turbulence}

\author{B. N. Rogers }

\affiliation{Dartmouth College, Hanover, NH 03825}

\affiliation{Center for Integrated Computation and Analysis of Reconnection and
Turbulence}
\begin{abstract}
Thin current sheets in systems of large size that exceed a critical
value of the Lundquist number are unstable to a super-Alfvénic tearing
instability, referred to hereafter as the plasmoid instability. The
scaling of the growth rate of the most rapidly growing plasmoid instability
with respect to the Lundquist number is shown to follow from the classical
dispersion relation for tearing modes. As a result of this instability,
the system realizes a nonlinear reconnection rate that appears to
be weakly dependent on the Lundquist number, and larger than the Sweet-Parker
rate by nearly an order of magnitude (for the range of Lundquist numbers
considered). This regime of fast reconnection is realizable in a dynamic
and highly unstable thin current sheet, without requiring the current
sheet to be turbulent. 
\end{abstract}
\maketitle
The problem of fast reconnection in high-Lundquist-number plasmas
has attracted a great deal of attention since the inception of the
classical Sweet-Parker \cite{Sweet1958,Parker1963} and Petschek \cite{Petschek1964}
models. In the Sweet-Parker model, the reconnection layer has the
structure of Y-points, with a length of the order of the system size
$L$, and a width given by $\delta_{SP}=L/S_{L}^{1/2}$, where $S_{L}$
is the Lundquist number based on the system size. The steady-state
reconnection time scale $\tau_{SP}$ is a strong function of $S_{L}$,
and increases as $S_{L}^{1/2}$. For weakly collisional systems such
as the solar corona, the Lundquist number is typically very large
$(\sim10^{12}-10^{14})$ and hence, the time scale $\tau_{SP}$ is
of the order of years, which is much too long to account for fast
events such as solar flares. A similar difficulty appears when the
Sweet-Parker model (equivalently, the Kadomtsev model \cite{Kadomtsev1976})
is applied to the problem of sawtooth crashes in high-temperature
tokamaks, where the predictions of the model for the sawtooth crash
time is significantly larger than that observed. These inadequacies
of the Sweet-Parker model has led to the point of view that within
the framework of the resistive MHD model, it is not possible to realize
fast reconnection for high-$S$ plasmas. 

In this Letter, we demonstrate that this widely accepted point of
view is not correct for large systems, characterized by high Lundquist
number ($S_{L}>3\times10^{4}$). By large systems, we mean systems
for which the aspect ratio of the thin current sheet, $\delta_{SP}/L$,
is much smaller than unity. In such systems, beyond a critical value
of $S_{L}$, there appears to be three qualitatively distinct phases
during the nonlinear evolution of reconnection. In the early nonlinear
regime, the system evolves to a quasi-steady but transient state with
the characteristics of Sweet-Parker reconnection --- an extended thin
current sheet in which reconnection occurs on the time-scale $\tau_{SP}$.
This quasi-steady regime is short-lived, and is followed by a second
phase in which a rapid secondary instability (referred to hereafter
as the plasmoid instability) occurs and produces plasmoids copiously.
Although the presence of this instability has been recognized for
some time \cite{BuchlinGV2005,LeeF1986,Biskamp1986,YanLP1992,ShibataT2001,Lapenta2008},
it is only relatively recently that the precise scaling properties
of the most rapidly growing plasmoid instability have been established\cite{LoureiroSC2007,NiBY2008,SamtaneyLUSC2009}.
We demonstrate here that the key properties of this instability can
be obtained as a simple extension of the classical dispersion relation
of tearing modes \cite{CoppiGPRR1976}, applied to a Sweet-Parker
current layer of width $\delta_{SP}$. The instability is followed
by a third nonlinear phase in which the system exhibits rapid and
impulsive reconnection mediated by a hierarchy of current sheets \cite{ShibataT2001}
with widths that are much smaller than $\delta_{SP}$, with a Lundquist
number dependency that decays faster than $S_{L}^{-1/2}$. When averaged
over time, reconnection in the third nonlinear regime proceeds at
a rate much faster than Sweet-Parker. Unlike the Sweet-Parker reconnection
rate, which decays as $S_{L}^{-1/2}$ with increasing $S_{L}$, the
reconnection rate in the third nonlinear phase appears to have a weak
dependence on $S_{L}$.

\emph{Linear Plasmoid Instability.} To fix ideas, consider a Harris
sheet of width a with the equilibrium magnetic field $\mathbf{B}=B_{0}\tanh(z/a)\mathbf{\hat{x}}$,
which reverses sign at the $z=0$ surface. According to classical
linear tearing instability theory, the tearing mode growth rate for
the \textquotedblleft{}constant-$\psi$\textquotedblright{} and \textquotedblleft{}non-
constant-$\psi$\textquotedblright{} modes are as follows, respectively
\cite{CoppiGPRR1976}:\begin{equation}
\gamma\tau_{A}\sim\left\{ \begin{array}{l}
S^{-3/5}\left(ka\right)^{-2/5}\left(1-k^{2}a^{2}\right)^{2/5},\, kaS^{1/4}\gg1\\
S^{-1/3}\left(ka\right)^{2/3},\, kaS^{1/4}\ll1,\, ka\ll1\end{array}\right.\label{eq:1}\end{equation}

Here $S=\tau_{R}/\tau_{A}$, where $\tau_{A}=a/V_{A}=a(4\pi\rho)^{1/2}/B_{0}$
is the Alfvén time, $\tau_{R}=4\pi a^{2}/\eta c^{2}$ is the resistive
diffusion time, $\rho$ is the mass density, $\eta$ is the resistivity
of the plasma, assumed to be a constant, and $c$ is the speed of
light. Note that the Lundquist number $S_{L}$, introduced earlier,
is related to $S$ through the relation, $S_{L}=S/\epsilon$, where
$\epsilon=a/L$. According to the dispersion relation \ref{eq:1},
the transition from \textquotedblleft{}constant-$\psi$\textquotedblright{}
to \textquotedblleft{}non-constant-$\psi$\textquotedblright{} modes
occurs at $kaS^{1/4}\sim1$, with a peak growth rate that scales as
$\gamma_{max}\tau_{A}\sim S^{-1/2}$. Since both branches of the tearing
mode dispersion relation \ref{eq:1} yield instabilities that grow
at a rate proportional to $S$ raised to a fractional \emph{negative}
exponent, it may seem surprising upon first glance that there exists
a super-Alfvénic tearing instability that actually grows at a rate
proportional to $S_{L}$ raised to a fractional\emph{ positive} exponent.
The key point is that unlike the Harris sheet, which is characterized
by an equilibrium width $a$ that is independent of the resistivity,
the Sweet-Parker current sheet is characterized by a width $\delta_{SP}$
that has a strong dependence on the resistivity. To connect with the
results obtained in \cite{LoureiroSC2007}, we introduce the dimensionless
wave number $\kappa=KL=ka/\epsilon$, and the dimensionless growth
rate $\gamma_{L}=\gamma\tau_{A}/\epsilon$. In terms of dimensionless
parameters $\kappa$, $\epsilon$, and $S_{L}$, the maximum growth
rate is given by $\gamma_{L,max}\sim\epsilon^{-2/3}\kappa^{2/3}S_{L}^{-1/3}$,
and it occurs at the wave number $\kappa_{max}\sim\epsilon^{-5/4}S_{L}^{-1/4}$.
For a Sweet-Parker current sheet, which obeys $\epsilon\sim S_{L}^{-1/2}$,
we then obtain $\kappa_{max}\sim S_{L}^{3/8}$, and the maximum growth
rate $\gamma_{L,max}\sim S_{L}^{1/4}$. Note that the number of plasmoids
generated in the linear regime of this instability scales as $\kappa_{max}$.
These scaling relations are completely consistent with the results
obtained for the fastest growing instability in \cite{LoureiroSC2007}. 

The rapidity of this instability is not limited to Sweet-Parker current
sheets, for we can easily extend the results obtained above to current
sheets that may be narrower than Sweet-Parker current sheets (for
the same value of $S_{L}$). If we write $\epsilon\sim S_{L}^{-\alpha}$,
where $\alpha>0$, we obtain $\kappa_{max}\sim S_{L}^{(5\alpha-1)/4}$
and $\gamma_{L,max}\sim S_{L}^{(3\alpha-1)/2}$. It follows then that
in sheets for which $\alpha>1/2$, the plasmoid instability will grow
even more rapidly than in Sweet-Parker current sheets of the same
Lundquist number (for which $\alpha=1/2$).

\emph{Nonlinear Evolution of Extended Current Sheets.} Our nonlinear
study is carried in two-dimensional (2D), doubly periodic geometry
using fully compressible MHD equations (in dimensionless form). The
magnetic field is represented as $\mathbf{B}=\nabla\psi(x,z)\times\mathbf{\hat{y}}$
where $\psi$ is a flux function and $y$ is the ignorable coordinate,
$\mathbf{v}=(v_{x},v_{z})$ is the fluid velocity, and $p$ is the
scalar pressure, assumed to obey an isothermal equation of state,
$p=2\rho T$, where $\rho$ is the density, and $T$ is a constant
temperature. The relevant dynamical equations are \begin{equation}
\partial_{t}\rho+\nabla\cdot\left(\rho\mathbf{v}\right)=0,\label{eq:2}\end{equation}
\begin{equation}
\partial_{t}\rho+\nabla\cdot\left(\rho\mathbf{vv}\right)=-\nabla p-\nabla\psi\nabla^{2}\psi,\label{eq:3}\end{equation}
\begin{equation}
\partial_{t}\psi+\mathbf{v}\cdot\nabla\psi=\eta\nabla^{2}\psi.\label{eq:4}\end{equation}
Our numerical algorithm \cite{GuzdarDMHL1993} approximates spatial
derivatives by finite differences with a five-point stencil in each
direction, and time derivatives by a second-order accurate trapezoidal
leapfrog method. We use variable grids in order to resolve the sharp
spatial gradients in the reconnection layer. The initial condition
consists of four flux tubes, which exhibit the coalescence instability.
The initial equilibrium state is given by $\psi=A_{0}\sin\left(2\pi x/L\right)\sin\left(2\pi z/L\right)$.
In the results that follow, we take $L=\sqrt{2}$ and $A_{0}=L/2\pi$,
which produces a peak magnetic field of unit magnitude in equilibrium.
The initial density is taken to be approximately unity and the pressure
(calculated from the isothermal equation of state with $T=3$ throughout
all simulations) is approximately equal to $6$. Both density and
pressure profiles are assumed to have a weak spatial dependence in
order to satisfy the magnetostatic equilibrium condition.

\begin{figure}[t]
\begin{centering}
\includegraphics[scale=0.8]{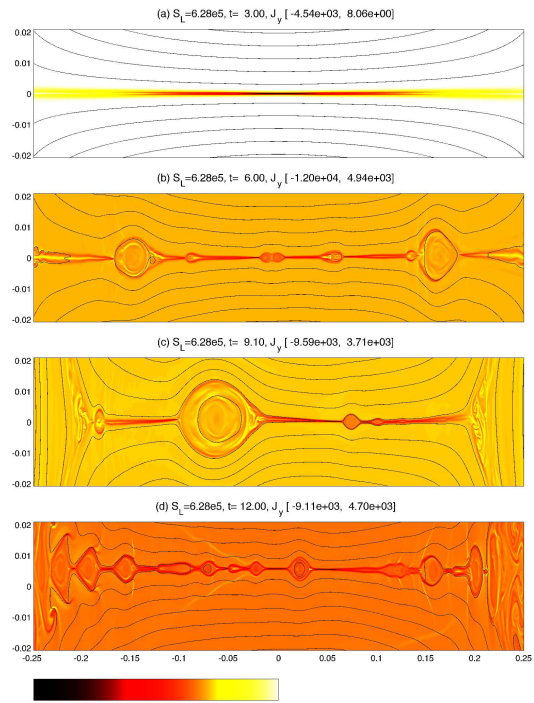}
\par\end{centering}

\caption{Time-sequence of the nonlinear evolution of the current density $J_{y}$
of a Sweet-Parker current sheet in a large system of Lundquist number
$S_{L}=6.28\times10^{5}$. The black lines represent surfaces of constant
$\psi$. \label{fig:1}}

\end{figure}

The flux tubes in equilibrium are unstable to the coalescence instability,
and produce an extended current sheet nonlinearly. Figure \ref{fig:1}
shows a time-sequence of the current density $J_{y}(x,z)$ for $S_{L}=6.28\times10^{5}$,
which exhibits the three stages of the nonlinear dynamics, discussed
above. (Here time is measured in units of $\tau_{AL}=L/V_{A}$.) For
convenience of visualization, we have rotated the current sheet by
45 degrees to make it horizontal. Frame (a) represents the first phase,
the so-called Sweet-Parker phase of the instability, which is characterized
by the formation of extended, system-size current sheets with the
structure of Y-points. Subsequently, this current sheet becomes spontaneously
unstable to the plasmoid instability, producing the second phase of
nonlinear evolution (frame (b)). This second phase is followed by
a third phase in which reconnection is typically impulsive (or bursty),
significantly faster than Sweet-Parker, and associated with the spontaneous
development and ejection of plasmoids in the outflow direction(s)
(frame (c)). As mentioned above, the third phase has multiple stages.
As the secondary instability develops, the initial Sweet-Parker current
sheet breaks up, producing islands that separate shorter, narrower
and more intense current sheets of width $\delta<\delta_{SP}$. These
intermediate current sheets become unstable to secondary tearing instabilities
of higher growth rate than those of the initial Sweet-Parker sheet,
and lead to a regime of faster reconnection (frame (d)). 

\begin{figure}[t]
\begin{centering}
\includegraphics[scale=0.5]{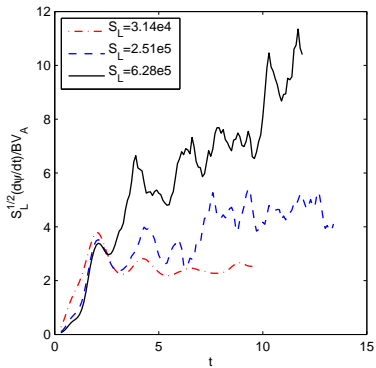}
\par\end{centering}

\caption{Time-evolution of the scaled reconnection rate for three values of
the Lundquist number. The black curve corresponds to the case shown
in Figure \ref{fig:1}. \label{fig:2}}

\end{figure}
\begin{figure}[t]
\begin{centering}
\includegraphics[scale=0.5]{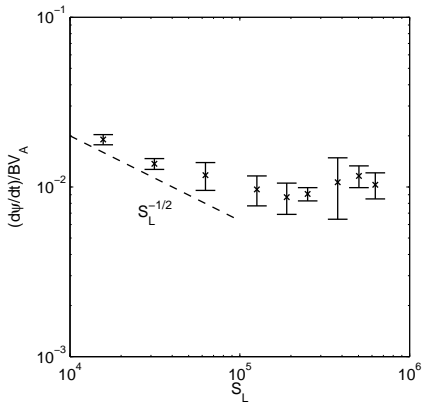}
\par\end{centering}

\caption{Averaged reconnection rate as a function of $S_{L}$ Below $S_{L}\simeq3\times10^{4}$,
the reconnection rate is Sweet-Parker and decays as $S_{L}^{-1/2}$,
shown by the dotted line. Above this threshold, the reconnection rate
becomes weakly dependent on $S_{L}$.\label{fig:3}}

\end{figure}

In Figure \ref{fig:2}, we show the time-history of the (normalized)
reconnection rate for three simulations, including the high-$S_{L}$
simulation (black plot) shown in Figure \ref{fig:1}. The reconnection
layer is along the x direction. We use a uniform grid along $x$.
The grid along $z$ is highly packed around $z=0$. Table \ref{tab:1}
summarizes the number of grid points as well as (the smallest) grid
size in each direction. (The Sweet-Parker thickness is included in
for reference.) The reconnection rate is calculated by taking the
time derivative of the maximum stream function in the reconnection
layer, normalized to the product of the upstream magnetic field and
the Alfvén speed, and multiplied by $S_{L}^{1/2}$. In Figure \ref{fig:2},
the black curve, which corresponds to $S_{L}=6.28\times10^{5}$, has
been smoothed over $0.4\tau_{A}$ to eliminate rapid variations in
a very dynamic current sheet. The linear instability of the initial
state produces a Sweet-Parker current sheet that persists until about
$t\sim3$. After this point in time, the secondary instability develops,
breaking up the long Sweet-Parker current sheet. (While the qualitative
features of the instability are those of the plasmoid instability
discussed above, it is difficult to test the precise scaling properties
of the linear instability during a nonlinear evolution.) There is
a rapid enhancement in the reconnection rate that remains impulsive
and fluctuates widely even as the rate tends to settles down to a
plateau, until about $t\sim9$. At this stage of the third nonlinear
phase, some of the small islands produced by the secondary instability
coalesce to form larger islands that are convected towards the boundaries.
(If the islands grow to large size but are constrained to stay fixed
at the center of the computational domain by reason of symmetries
imposed in the simulations, the third nonlinear phase may be short-lived,
and the reconnection rate may fall rapidly.) At about this point in
time, the extended current sheet shows yet another burst of secondary
tearing activity producing multiple plasmoids, and a consequent enhancement
in the reconnection rate, which at about $t\sim12$ attains an order
of magnitude higher than the Sweet-Parker rate at this value of $S_{L}$.
Due to insufficient spatial resolution, caused by the slow drift of
the current sheet away from the region where the grid points along
$z$ are clustered, we are not able to carry these simulations forward
longer in time. 

The plasmoid instability of Sweet-Parker sheets occurs after $S_{L}$
exceeds a critical value, determined numerically to be approximately
$3\times10^{4}$ in the present study. Like the black curve, the blue
dashed curve in Figure \ref{fig:2} corresponds to another value of
$S_{L}$ ($=2.51\times10^{5}$) above the threshold and shows generically
similar behavior, while the red dashed curve corresponds to a value
($=3.14\times10^{4}$) at about the threshold. We have identified
the plateau regions in these curves by inspection, and stated the
corresponding time-intervals in Table \ref{tab:1}. There is clearly
an element of arbitrariness in this identification, because the underlying
dynamics is not quasi-steady. In fact, the dynamics is generically
impulsive, strongly time-dependent, involving the growth and ejection
of multiple plasmoids in the presence, and complicated by the sloshing
of coalescing islands as well as the tilt and slow drift of the current
sheet away from the central region. The sloshing is most clearly seen
in the oscillations of the red curve, where the first peak corresponds
to the highest reconnection rate, after which the system settles down
to a Sweet-Parker-like plateau regime. A rough definition of the plateau
region is when the reconnection rate appears to fluctuate about a
relatively steady value. If the Lundquist number of the simulation
is above the threshold for the plasmoid instability, the plateau region
is typically characterized by the presence of multiple islands. 

Figure \ref{fig:3} shows a plot of the averaged reconnection rate
(normalized) as a function of $S_{L}$. The averaged reconnection
rate is calculated by averaging the smoothed reconnection rate curve
over the plateau region. The error bar is calculated from the standard
deviation of the data points in the plateau region. Our simulations
suggest that below a threshold, $S_{L}=3\times10^{4}$, the reconnection
rate exhibits Sweet-Parker scaling. Over this threshold, the averaged
reconnection rate exhibits a much weaker dependency on $S_{L}$. At
high values of $S_{L}$, the average reconnection rate exceeds the
Sweet-Parker rate by nearly an order of magnitude. We conjecture that
as the initial Sweet-Parker current sheet is broken up into sufficiently
short segments, and the effective Lundquist number of these segments
falls below the threshold of instability, the process will attain
a quasi-steady state, provided the reconnection dynamics is not quenched
by the erosion of the flux content of the system due to the cumulative
effects of resistive diffusion.

These results bear qualitative resemblance and are complementary to
those reported in a concurrent study \cite{DaughtonRAKYB2009}, where
2D, fully electromagnetic particle-in-cell simulations that include
a collision operator show a regime of fast reconnection brought about
by the onset of secondary tearing instabilities. Since the physical
models underlying the simulations presented here and in \cite{DaughtonRAKYB2009}
are quite different, it is not surprising that there are some important
differences between the two studies. The initial Sweet-parker current
sheet realized in \cite{DaughtonRAKYB2009} shows clear evidence of
the plasmoid instability, and the width of subsequent intermediate
current sheet segments becomes much narrower as time progresses, as
in the present study. Whereas this narrowing is controlled entirely
by the value of the Lundquist number and can continue without the
intervention of kinetic effects in the present study, kinetic effects
inevitably intervene in \cite{DaughtonRAKYB2009} when the current
sheet becomes of the order of ion skin depth. For systems of moderate
size (of the order of $200d_{i}$ or less), when $\delta\le d_{i}$,
the simulations in \cite{DaughtonRAKYB2009} show the onset of fast
reconnection, consistent with the predictions of earlier Hall MHD
models \cite{MaB1996a,Bhattacharjee2004,CassakSD2005,SimakovC2008},
and verified by experiment \cite{YamadaRJBGKK2006}. However, for
systems of larger size, the transition to fast reconnection appears
to occur at values of Lundquist number that are somewhat smaller than
the Hall MHD prediction \cite{MaB1996a,Bhattacharjee2004,CassakSD2005,SimakovC2008}
due to the intervention of the plasmoid instability. 

In summary, we have demonstrated that within the framework of the
resistive MHD model, thin current sheets in systems of large size
that exceed a critical value of the Lundquist number are unstable
to a super-Alfvénic tearing instability, referred to here as the plasmoid
instability. If the aspect ratio of the current sheet is written $\epsilon\sim S_{L}^{-\alpha}$,
where $\alpha>0$, the fastest growing instability is characterized
by a wave number $\kappa_{max}\sim S_{L}^{(5\alpha-1)/4}$ and a growth
rate $\gamma_{L,max}\sim S_{L}^{(3\alpha-1)/2}$. In the special case
of a Sweet-Parker current sheet for which $\alpha=1/2$, we obtain
the instability discussed in \cite{LoureiroSC2007}. As a result of
this class of instabilities, the system realizes a nonlinear reconnection
rate that appears to be weakly dependent on the Lundquist number,
and larger than the Sweet-Parker rate by an order of magnitude (for
the range of Lundquist numbers considered in the present study). We
note that this weak dependency on the Lundquist number is realized
in a thin current sheet that is violently unstable and dynamic, but
not turbulent. In this respect, our results are different from those
obtained from turbulent reconnection studies in 2D \cite{Lapenta2008,MatthaeusL1986,LoureiroUSCY2009}
as well as 3D \cite{LazarianV1999,KowalLVO2009}, which also report
reconnection rates that depend weakly on the resistivity. 

This research is supported by the Department of Energy under Grant
No. DE-FG02-07ER46372, by NASA under Grant Nos. NNX06AC19G and NNX09AJ86G,
and by NSF under Grant No. ATM-090315. Supercomputing support is provided
by the National Energy Research Supercomputing Center.

\begin{table}
\caption{Details of the spatial resolution and the plateau regime for the runs
shown in Figure \ref{fig:2}\label{tab:1}}

\begin{centering}
\begin{longtable}{ccccccc}
\hline
\hline 
$S$ & $\Delta x$ & $\Delta z$ & $N_{x}$ & $N_{z}$ & $S_{L}^{-1/2}$ & Plateau Regime \tabularnewline
\endhead
2.51e5 & 5e-4 & 9.4e-5 & 2001 & 1001 & 2e-3 & 6.9--13.0\tabularnewline
6.28e5 & 3.3e-4 & 3.9e-5 & 3001 & 1001 & 1.26e-3 & 6.9--11.6\tabularnewline
\hline
\hline
\endfoot
\hline 
3.14e4 & 6.7e-4 & 5e-4 & 1501 & 501 & 5.6e-3 & 2.9--8.9\tabularnewline
\end{longtable}
\par\end{centering}

\end{table}

\bibliographystyle{apsrev}

\begin{thebibliography}{25}
\expandafter\ifx\csname natexlab\endcsname\relax\def\natexlab#1{#1}\fi
\expandafter\ifx\csname bibnamefont\endcsname\relax
  \def\bibnamefont#1{#1}\fi
\expandafter\ifx\csname bibfnamefont\endcsname\relax
  \def\bibfnamefont#1{#1}\fi
\expandafter\ifx\csname citenamefont\endcsname\relax
  \def\citenamefont#1{#1}\fi
\expandafter\ifx\csname url\endcsname\relax
  \def\url#1{\texttt{#1}}\fi
\expandafter\ifx\csname urlprefix\endcsname\relax\def\urlprefix{URL }\fi
\providecommand{\bibinfo}[2]{#2}
\providecommand{\eprint}[2][]{\url{#2}}

\bibitem[{\citenamefont{Sweet}(1958)}]{Sweet1958}
\bibinfo{author}{\bibfnamefont{P.~A.} \bibnamefont{Sweet}},
  \bibinfo{journal}{Nuovo Cimento Suppl. Ser. X} \textbf{\bibinfo{volume}{8}},
  \bibinfo{pages}{188} (\bibinfo{year}{1958}).

\bibitem[{\citenamefont{Parker}(1963)}]{Parker1963}
\bibinfo{author}{\bibfnamefont{E.~N.} \bibnamefont{Parker}},
  \bibinfo{journal}{Astrophys. J. Suppl.} \textbf{\bibinfo{volume}{8}},
  \bibinfo{pages}{177} (\bibinfo{year}{1963}).

\bibitem[{\citenamefont{Petschek}(1964)}]{Petschek1964}
\bibinfo{author}{\bibfnamefont{H.~E.} \bibnamefont{Petschek}}, in
  \emph{\bibinfo{booktitle}{AAS/NASA Symposium on the Physics of Solar
  Flares}}, edited by \bibinfo{editor}{\bibfnamefont{W.~N.} \bibnamefont{Hess}}
  (\bibinfo{publisher}{NASA}, \bibinfo{address}{Washington, DC},
  \bibinfo{year}{1964}), p. \bibinfo{pages}{425}.

\bibitem[{\citenamefont{Kadomtsev}(1976)}]{Kadomtsev1976}
\bibinfo{author}{\bibfnamefont{B.~B.} \bibnamefont{Kadomtsev}},
  \bibinfo{journal}{Sov. J. Plasma Phys.} \textbf{\bibinfo{volume}{2}},
  \bibinfo{pages}{389} (\bibinfo{year}{1976}).

\bibitem[{\citenamefont{Buchlin et~al.}(2005)\citenamefont{Buchlin, Galtier,
  and Velli}}]{BuchlinGV2005}
\bibinfo{author}{\bibfnamefont{E.}~\bibnamefont{Buchlin}},
  \bibinfo{author}{\bibfnamefont{S.}~\bibnamefont{Galtier}}, \bibnamefont{and}
  \bibinfo{author}{\bibfnamefont{M.}~\bibnamefont{Velli}},
  \bibinfo{journal}{Astron. and Astrophys.} \textbf{\bibinfo{volume}{436}},
  \bibinfo{pages}{355} (\bibinfo{year}{2005}).

\bibitem[{\citenamefont{Lee and Fu}(1986)}]{LeeF1986}
\bibinfo{author}{\bibfnamefont{L.~C.} \bibnamefont{Lee}} \bibnamefont{and}
  \bibinfo{author}{\bibfnamefont{Z.~F.} \bibnamefont{Fu}}, \bibinfo{journal}{J.
  Geophys. Res.} \textbf{\bibinfo{volume}{91}}, \bibinfo{pages}{6807}
  (\bibinfo{year}{1986}).

\bibitem[{\citenamefont{Biskamp}(1986)}]{Biskamp1986}
\bibinfo{author}{\bibfnamefont{D.}~\bibnamefont{Biskamp}},
  \bibinfo{journal}{Phys. Fluids} \textbf{\bibinfo{volume}{29}},
  \bibinfo{pages}{1520} (\bibinfo{year}{1986}).

\bibitem[{\citenamefont{Yan et~al.}(1992)\citenamefont{Yan, Lee, and
  Priest}}]{YanLP1992}
\bibinfo{author}{\bibfnamefont{M.}~\bibnamefont{Yan}},
  \bibinfo{author}{\bibfnamefont{H.~C.} \bibnamefont{Lee}}, \bibnamefont{and}
  \bibinfo{author}{\bibfnamefont{E.~R.} \bibnamefont{Priest}},
  \bibinfo{journal}{Journal of Geophysical Research}
  \textbf{\bibinfo{volume}{97}}, \bibinfo{pages}{8277} (\bibinfo{year}{1992}).

\bibitem[{\citenamefont{Shibata and Tanuma}(2001)}]{ShibataT2001}
\bibinfo{author}{\bibfnamefont{K.}~\bibnamefont{Shibata}} \bibnamefont{and}
  \bibinfo{author}{\bibfnamefont{S.}~\bibnamefont{Tanuma}},
  \bibinfo{journal}{Earth Planets Space} \textbf{\bibinfo{volume}{53}},
  \bibinfo{pages}{473} (\bibinfo{year}{2001}).
\bibitem[{\citenamefont{Lapenta}(2008)}]{Lapenta2008}
\bibinfo{author}{\bibfnamefont{G.}~\bibnamefont{Lapenta}},
  \bibinfo{journal}{Phys. Rev. Lett.} \textbf{\bibinfo{volume}{100}},
  \bibinfo{pages}{235001} (\bibinfo{year}{2008}).

\bibitem[{\citenamefont{Loureiro et~al.}(2007)\citenamefont{Loureiro,
  Schekochihin, and Cowley}}]{LoureiroSC2007}
\bibinfo{author}{\bibfnamefont{N.~F.} \bibnamefont{Loureiro}},
  \bibinfo{author}{\bibfnamefont{A.~A.} \bibnamefont{Schekochihin}},
  \bibnamefont{and} \bibinfo{author}{\bibfnamefont{S.~C.}
  \bibnamefont{Cowley}}, \bibinfo{journal}{Phys. Plasmas}
  \textbf{\bibinfo{volume}{14}}, \bibinfo{pages}{100703}
  (\bibinfo{year}{2007}).

\bibitem[{\citenamefont{Ni et~al.}(2008)\citenamefont{Ni, Bhattacharjee, and
  Yang}}]{NiBY2008}
\bibinfo{author}{\bibfnamefont{L.}~\bibnamefont{Ni}},
  \bibinfo{author}{\bibfnamefont{A.}~\bibnamefont{Bhattacharjee}},
  \bibnamefont{and} \bibinfo{author}{\bibfnamefont{H.}~\bibnamefont{Yang}},
  \bibinfo{journal}{Bull. Am. Phys. Soc.} \textbf{\bibinfo{volume}{53}},
  \bibinfo{pages}{94} (\bibinfo{year}{2008}).

\bibitem[{\citenamefont{Samtaney et~al.}(2009)\citenamefont{Samtaney, Loureiro,
  Uzdensky, Schekochihin, and Cowley}}]{SamtaneyLUSC2009}
\bibinfo{author}{\bibfnamefont{R.}~\bibnamefont{Samtaney}},
  \bibinfo{author}{\bibfnamefont{N.~F.} \bibnamefont{Loureiro}},
  \bibinfo{author}{\bibfnamefont{D.~A.} \bibnamefont{Uzdensky}},
  \bibinfo{author}{\bibfnamefont{A.~A.} \bibnamefont{Schekochihin}},
  \bibnamefont{and} \bibinfo{author}{\bibfnamefont{S.~C.}
  \bibnamefont{Cowley}}, \bibinfo{journal}{submitted to PRL}
  (\bibinfo{year}{2009}).

\bibitem[{\citenamefont{Coppi et~al.}(1976)\citenamefont{Coppi, Galvao, Pellat,
  Rosenbluth, and Rutherford}}]{CoppiGPRR1976}
\bibinfo{author}{\bibfnamefont{B.}~\bibnamefont{Coppi}},
  \bibinfo{author}{\bibfnamefont{E.}~\bibnamefont{Galvao}},
  \bibinfo{author}{\bibfnamefont{R.}~\bibnamefont{Pellat}},
  \bibinfo{author}{\bibfnamefont{M.~N.} \bibnamefont{Rosenbluth}},
  \bibnamefont{and} \bibinfo{author}{\bibfnamefont{P.~H.}
  \bibnamefont{Rutherford}}, \bibinfo{journal}{Sov. J. Plasma Phys.}
  \textbf{\bibinfo{volume}{2}}, \bibinfo{pages}{533} (\bibinfo{year}{1976}).

\bibitem[{\citenamefont{Guzdar et~al.}(1993)\citenamefont{Guzdar, Drake,
  McCarthy, Hassam, and Liu}}]{GuzdarDMHL1993}
\bibinfo{author}{\bibfnamefont{P.~N.} \bibnamefont{Guzdar}},
  \bibinfo{author}{\bibfnamefont{J.~F.} \bibnamefont{Drake}},
  \bibinfo{author}{\bibfnamefont{D.}~\bibnamefont{McCarthy}},
  \bibinfo{author}{\bibfnamefont{A.~B.} \bibnamefont{Hassam}},
  \bibnamefont{and} \bibinfo{author}{\bibfnamefont{C.~S.} \bibnamefont{Liu}},
  \bibinfo{journal}{Phys. Fluids B} \textbf{\bibinfo{volume}{5}},
  \bibinfo{pages}{3712} (\bibinfo{year}{1993}).

\bibitem[{\citenamefont{Daughton et~al.}(2009)\citenamefont{Daughton,
  Roytershteyn, Albright, Karimabadi, Yin, and Bowers}}]{DaughtonRAKYB2009}
\bibinfo{author}{\bibfnamefont{W.}~\bibnamefont{Daughton}},
  \bibinfo{author}{\bibfnamefont{V.}~\bibnamefont{Roytershteyn}},
  \bibinfo{author}{\bibfnamefont{B.~J.} \bibnamefont{Albright}},
  \bibinfo{author}{\bibfnamefont{H.}~\bibnamefont{Karimabadi}},
  \bibinfo{author}{\bibfnamefont{L.}~\bibnamefont{Yin}}, \bibnamefont{and}
  \bibinfo{author}{\bibfnamefont{K.~J.} \bibnamefont{Bowers}},
  \bibinfo{journal}{Phys. Rev. Lett.} \textbf{\bibinfo{volume}{103}},
  \bibinfo{pages}{065004} (\bibinfo{year}{2009}).

\bibitem[{\citenamefont{Ma and Bhattacharjee}(1996)}]{MaB1996a}
\bibinfo{author}{\bibfnamefont{Z.~W.} \bibnamefont{Ma}} \bibnamefont{and}
  \bibinfo{author}{\bibfnamefont{A.}~\bibnamefont{Bhattacharjee}},
  \bibinfo{journal}{Geophys. Res. Lett.} \textbf{\bibinfo{volume}{23}},
  \bibinfo{pages}{1673} (\bibinfo{year}{1996}).

\bibitem[{\citenamefont{Bhattacharjee}(2004)}]{Bhattacharjee2004}
\bibinfo{author}{\bibfnamefont{A.}~\bibnamefont{Bhattacharjee}},
  \bibinfo{journal}{Annu. Rev. Astron. Astrophys.}
  \textbf{\bibinfo{volume}{42}}, \bibinfo{pages}{365} (\bibinfo{year}{2004}).

\bibitem[{\citenamefont{Cassak et~al.}(2005)\citenamefont{Cassak, Shay, and
  Drake}}]{CassakSD2005}
\bibinfo{author}{\bibfnamefont{P.~A.} \bibnamefont{Cassak}},
  \bibinfo{author}{\bibfnamefont{M.~A.} \bibnamefont{Shay}}, \bibnamefont{and}
  \bibinfo{author}{\bibfnamefont{J.~F.} \bibnamefont{Drake}},
  \bibinfo{journal}{Phys. Rev. Lett.} \textbf{\bibinfo{volume}{95}},
  \bibinfo{pages}{235002} (\bibinfo{year}{2005}).

\bibitem[{\citenamefont{Simakov and Chac\'on}(2008)}]{SimakovC2008}
\bibinfo{author}{\bibfnamefont{A.~N.} \bibnamefont{Simakov}} \bibnamefont{and}
  \bibinfo{author}{\bibfnamefont{L.}~\bibnamefont{Chac\'on}},
  \bibinfo{journal}{Phys. Rev. Lett.} \textbf{\bibinfo{volume}{101}},
  \bibinfo{pages}{105003} (\bibinfo{year}{2008}).

\bibitem[{\citenamefont{Yamada et~al.}(2006)\citenamefont{Yamada, Ren, Ji,
  Breslau, Gerhardt, Kulsrud, and Kuritsyn}}]{YamadaRJBGKK2006}
\bibinfo{author}{\bibfnamefont{M.}~\bibnamefont{Yamada}},
  \bibinfo{author}{\bibfnamefont{Y.}~\bibnamefont{Ren}},
  \bibinfo{author}{\bibfnamefont{H.}~\bibnamefont{Ji}},
  \bibinfo{author}{\bibfnamefont{J.}~\bibnamefont{Breslau}},
  \bibinfo{author}{\bibfnamefont{S.}~\bibnamefont{Gerhardt}},
  \bibinfo{author}{\bibfnamefont{R.}~\bibnamefont{Kulsrud}}, \bibnamefont{and}
  \bibinfo{author}{\bibfnamefont{A.}~\bibnamefont{Kuritsyn}},
  \bibinfo{journal}{Phys. Plasmas} \textbf{\bibinfo{volume}{13}},
  \bibinfo{pages}{052119} (\bibinfo{year}{2006}).

\bibitem[{\citenamefont{Matthaeus and Lamkin}(1986)}]{MatthaeusL1986}
\bibinfo{author}{\bibfnamefont{W.~H.} \bibnamefont{Matthaeus}}
  \bibnamefont{and} \bibinfo{author}{\bibfnamefont{S.~L.}
  \bibnamefont{Lamkin}}, \bibinfo{journal}{Phys. Fluids}
  \textbf{\bibinfo{volume}{29}}, \bibinfo{pages}{2513} (\bibinfo{year}{1986}).

\bibitem[{\citenamefont{Loureiro et~al.}(2009)\citenamefont{Loureiro, Uzdensky,
  Schekochihin, Cowley, and Yousef}}]{LoureiroUSCY2009}
\bibinfo{author}{\bibfnamefont{N.~F.} \bibnamefont{Loureiro}},
  \bibinfo{author}{\bibfnamefont{D.~A.} \bibnamefont{Uzdensky}},
  \bibinfo{author}{\bibfnamefont{A.~A.} \bibnamefont{Schekochihin}},
  \bibinfo{author}{\bibfnamefont{S.~C.} \bibnamefont{Cowley}},
  \bibnamefont{and} \bibinfo{author}{\bibfnamefont{T.~A.}
  \bibnamefont{Yousef}}, \bibinfo{journal}{submitted to PRL}
  (\bibinfo{year}{2009}).

\bibitem[{\citenamefont{Lazarian and Vishniac}(1999)}]{LazarianV1999}
\bibinfo{author}{\bibfnamefont{A.}~\bibnamefont{Lazarian}} \bibnamefont{and}
  \bibinfo{author}{\bibfnamefont{E.~T.} \bibnamefont{Vishniac}},
  \bibinfo{journal}{Astrophys. J.} \textbf{\bibinfo{volume}{517}},
  \bibinfo{pages}{700} (\bibinfo{year}{1999}).

\bibitem[{\citenamefont{Kowal et~al.}(2009)\citenamefont{Kowal, Lazarian,
  Vishniac, and Otmianowska-Mazur}}]{KowalLVO2009}
\bibinfo{author}{\bibfnamefont{G.}~\bibnamefont{Kowal}},
  \bibinfo{author}{\bibfnamefont{A.}~\bibnamefont{Lazarian}},
  \bibinfo{author}{\bibfnamefont{E.~T.} \bibnamefont{Vishniac}},
  \bibnamefont{and}
  \bibinfo{author}{\bibfnamefont{K.}~\bibnamefont{Otmianowska-Mazur}},
  \bibinfo{journal}{Astrophys. J.} \textbf{\bibinfo{volume}{700}},
  \bibinfo{pages}{63} (\bibinfo{year}{2009}).

\end{thebibliography}

\end{document}